\documentclass[conference,10pt]{IEEEtran}

\ifCLASSINFOpdf

\fi

\usepackage[subrefformat=parens,labelformat=parens,caption=false,font=footnotesize]{subfig}
\usepackage{array}
\usepackage{amsmath}
\usepackage{breqn}
\usepackage[nolist,nohyperlinks]{acronym}
\usepackage{graphicx,wrapfig,lipsum}
\usepackage{rotating}
\usepackage{cleveref}
\usepackage{epstopdf}
\usepackage{balance}
\usepackage{amsmath, amssymb, amsthm}
\usepackage{stmaryrd}
\usepackage{tikz}
\usepackage{verbatim}
\usepackage[short]{optidef}
\usepackage{url}
\usepackage[utf8]{inputenc}
\usepackage[english]{babel}

\usepackage{multicol}
\usepackage{xr-hyper}

\usepackage{scalerel}
\usepackage{multicol}

\usepackage[noadjust]{cite}
\usepackage[normalem]{ulem}
\usepackage{color}
\usepackage{dsfont}
\usepackage{bm}
\usepackage[short]{optidef}
\usepackage{diagbox}
\usepackage{enumitem}
\usepackage{slashbox}
\usepackage[ruled]{algorithm2e}
\usepackage{multirow}
\usepackage[T1]{}
\usepackage{color, colortbl}
\usepackage{babel}
\usepackage{verbatim}
\usepackage{textcomp}
\usepackage{tabulary}
\usepackage{lipsum}
\usepackage{caption}
\definecolor{Gray}{gray}{0.95}
\definecolor{LightCyan}{rgb}{0.8,0.85,1}
\definecolor{LightBlue}{rgb}{0.6,0.6,1}
\usepackage[font=small]{caption}

\setlist{nosep}
\newcommand\blfootnote[1]{%
  \begingroup
  \renewcommand\thefootnote{}\footnote{#1}%
  \addtocounter{footnote}{-1}%
  \endgroup
}
\usepackage{mathtools}

\begin{document}

\title{Throughput and Coverage Trade-Off in Integrated Terrestrial and Non-Terrestrial Networks: an Optimization Framework}

\author{Henri Alam$^{\dag\ddag}$, Antonio De Domenico$^\dag$, David L\'{o}pez-P\'{e}rez$^\bigstar$, Florian Kaltenberger$^\ddag$  \\
 \small{$^\dag$Huawei Technologies, Paris Research Center, 20 quai du Point du Jour, Boulogne Billancourt, France.} \\
 \small{$^\ddag$EURECOM, 2229 route des Cretes, 06904 Sophia Antipolis Cedex, France.} \\
 \small{$^\bigstar$Universitat Politècnica de València, Spain.}}
\maketitle
\vspace{-1.7cm}
\thispagestyle{empty}

\begin{abstract}
In past years, 
non-terrestrial networks (NTNs) have emerged as a viable solution for providing ubiquitous connectivity for future wireless networks due to their ability to reach large geographical areas.
However, the efficient integration and operation of an NTN with a classic terrestrial network (TN) is challenging due the large amount of parameters to tune.
In this paper, 
we consider the downlink scenario of an integrated TN-NTN transmitting over the S band, 
comprised of low-earth orbit (LEO) satellites overlapping a large-scale ground cellular network.
We propose a new resource management framework to optimize the user equipment (UE) performance by properly controlling the spectrum allocation, the UE association and the transmit power of ground base stations (BSs) and satellites.
Our study reveals that, in rural scenarios, NTNs, 
combined with the proposed radio resource management framework, 
reduce the number of UEs that are out of coverage, highlighting the important role of NTNs in providing ubiquitous connectivity, and greatly improve the overall capacity of the network. 
Specifically, our solution leads to more than $200\%$ gain in terms of mean data rate with respect to a network without satellites and a standard integrated TN-NTN when the resource allocation setting follows 3GPP recommendation. \blfootnote{This research was supported  by the Generalitat Valenciana, Spain, through the  CIDEGENT PlaGenT,  Grant CIDEXG/2022/17,  Project iTENTE.}

\end{abstract}

\section{Introduction}
\label{sec:intro}
In the midst of an era witnessing fast development of cellular communications, 
the demand for high-data-rate connectivity has soared. 
This has resulted in more stringent requirements on providing high capacity and guaranteeing ubiquitous connectivity for the network.
The usage of heterogeneous networks (HetNets) has proven to be an appealing solution in a bid to answer those demands \cite{Damnjanovic_2011}. 
Indeed, by creating a multi-tier architecture of the network, 
its inherent flexibility allows an effective data offloading, 
which in turn leads to higher capacity and better coverage throughout the network.
Recently, non-terrestrial networks (NTNs) have emerged as a viable solution to complement the terrestrial network (TN), 
and ensure that uncovered geographical areas can be served \cite{Ahmmed_2022}. 
An NTN is a network where aerial vehicles such as drones (i.e. UAVs), high-altitude platform station (HAPS) or satellites act as a relay node or a base station (BS) to serve the user equipment (UE) in the network.
The intrinsic benefit of NTNs is their ability to provide coverage for wide areas,  
reaching geographical locations where it would have been expensive or difficult to deploy macro BSs (MBS).
Among the different deployment options, 
it seems that the low-earth orbit satellites will spearhead the process of achieving high-capacity connectivity from space \cite{Giordani_2021,Rinaldi_2020}. 
The low-earth orbit (LEO) satellite is a non-geostationary satellite that orbits at an altitude between $200$ and $2000$ km.
Its shorter distance to Earth means that, 
compared to other satellite architectures, 
there will be a better signal strength and lower latency, less energy needed for launching and less power required for the transmission of the signal from/to the satellite.
Taking all this into account, 
the concept of an integrated TN-NTN may be the way forward to ensure efficient services for terrestrial and aerial UEs \cite{Benzaghta_2022}.
In most practical networks, 
each UE is associated to the BS which provides the highest reference signal received power (RSRP) in the network. 
This association policy has its limits since it does not account for the fluctuating traffic demands of UEs, 
and can lead to poor load balancing and thus performance.
A better performing UE association policy should take into account, 
not only the strength and/or quality of the UE signal, 
but also the load on each cell.
Although load balancing has been well studied in the cellular literature,
most of the related work does not consider NTNs. 
The most advanced analysis in this front \cite{Benzaghta_2022} has recently studied an integrated TN-NTN deployed in an urban area,
and has revealed that offloading part of the traffic to LEO satellites would not only improve the overall signal quality of the network, 
but also reduce outages.
However, optimal operation points are not derived. 
From a load balancing perspective, 
the common approach is to build a framework which maximizes a selected utility function using a pricing based association strategy \cite{Ye_2013, Shen_2014}.
In \cite{Di_2019} and \cite{Zhang_2022_journal}, 
following such approach, 
the authors study the uplink performance of an integrated TN-NTN where the LEO satellites are used to provide backhaul to the ground BSs.
The objective in both papers is to maximize the uplink sum data rate satisfying backhaul capacity constraints, while \cite{Zhang_2022_journal} also considers minimal rate constraints.
\cite{Di_2019} optimizes the user association and power allocation through matching algorithms.
In \cite{Zhang_2022_journal}, the authors also consider the split of the bandwidth between the fronthaul and backhaul link as a variable.

\noindent In this paper, {we extend the load balancing literature,} 
deriving for the first time the optimal radio resource management --in terms of joint bandwidth split, UE association, and power control-- in an integrated TN-NTN.
The objective is to improve the overall capacity of a large-scale rural network,
while providing coverage guarantees to all UEs,
by dynamically tailoring the complementary capabilities of both network tiers.
Importantly, 
our results show that the developed framework improves the mean data rate by more than $200$ $\%$ with respect to a network without satellites and a standard integrated TN-NTN with a resource allocation setting that follows the 3GPP recommendations \cite{3GPPTR36.814,3GPPTR38.821}.  


\section{System Model \& Problem Formulation}
\label{sec:SysModelProblemFormulation}

In this section, we present the system model and the problem formulation. 

\subsection{System Model}
 
We consider a downlink cellular network consisting of $M$ macro BSs and $N$ LEO satellites, 
all serving $K$ UEs that are deployed in a rural area.
We denote as $W$ the total bandwidth of the system, 
which the mobile network operator shares between the terrestrial and non-terrestrial tier.
In our study, 
we suppose that such bandwidth $W$ is operated over the S band, 
i.e. around $2$ GHz, 
and that macro and satellite BSs use orthogonal fractions of it.
In the remainder of the paper, 
we will denote by $\mathcal{T}$ (resp. $\mathcal{S}$) the set of terrestrial (resp. non-terrestrial) BSs. 
Moreover, let $\mathcal{U} =  \{1, \cdots, i, \cdots,K \}$ be the set of UEs 
and $\mathcal{B} = \mathcal{T} \cup \mathcal{S} = \{ 1, \cdots, j, \cdots, M+N \} $ the set of all BSs.
With respect to the channel model, 
the large-scale channel gain between a macro BS $j$ and a UE $i$ is calculated as follows:
\begin{equation}\label{channel_terrestrial}
     \beta_{ij}  = G_{T_X} \cdot PL_{ij} \cdot SF_{ij} ,
\end{equation}
where $G_{T_X}$ is the transmit antenna gain, 
$PL_{ij}$ is the path loss, 
and $SF_{ij}$ is the shadow fading.
On the contrary, 
if a satellite BS $j$ serves a UE $i$, 
then the large-scale channel gain is the following \cite{3GPPTR38.811}:
\begin{equation}\label{channel_satellite}
     \beta_{ij}  = G_{T_X} \cdot PL_{ij} \cdot SF_{ij} \cdot CL \cdot PL_s
\end{equation}
where $CL$ is the clutter loss, 
i.e. an attenuation caused by buildings and vegetation in the vicinity of the UE, 
and $PL_s$ is the scintillation loss (rapid variations in the amplitude and phase of the signal due to the structure of the ionosphere).
Considering that a UE will only be served by a macro BS or a satellite, 
and that the terrestrial and non-terrestrial tiers do not interfere each other due to the orthogonal bandwidth allocation,
we can compute the signal-to-interference-plus-noise ratio (SINR) for each UE $i$ as follows:
\begin{equation}\label{SINR}
    \gamma_{ij}  = \frac{ \beta_{ij} p_j}{ \sum\limits_{\substack{ j^' \in \mathcal{I}_j}} \beta_{ij^'} p_{j^'} + \sigma^2},
\end{equation}
where $p_j$ is the transmit power allocated per resource element (RE) at BS $j$, 
$\mathcal{I}_j$ is the set of BSs that are interfering with serving BS $j$,
and $\sigma^2$ is the noise power.
Thereafter, 
assuming that BS $j$ equally shares its available bandwidth $W_j$ among its $k_j$ served UEs, 
the average data rate for the UE $i$ connected to BS $j$ can be computed as:
\begin{equation}\label{Shannon_data_rate}
    R_{ij}  = \frac{W_j}{k_j} \log_2(1 + \gamma_{ij}).
\end{equation}

 \subsection{Problem Formulation}

Since we want to ensure a proportionally fair resource allocation,
our goal is to optimize the sum of the log-throughput (SLT) across all UEs in the network.
To achieve this goal, 
we want to find the optimal bandwidth split between the non-terrestrial and terrestrial tiers of the network.
Taking this into account, 
we introduce $\varepsilon$ as the share of the bandwidth allocated to the LEO satellites. 
Thus, the bandwidth $W_j$ of the BS $j$  can be computed as $W\varepsilon$ if it is a satellite or as $W(1-\varepsilon)$ if it is a macro BS.
Let us also define a binary variable $x_{i,j}$ which is equal to $1$ if UE $i$ is associated to the BS $j$,
and $0$ otherwise.
Our aim is then to optimize the UE-BS association, the transmit power allocation at each BS as well as the bandwidth allocation to each tier to maximize the SLT of the network.
This can be written as follows:
\begin{maxi!}|s|[2]
{\text{X},\text{p},\text{k}, \varepsilon }{\sum\limits_{i \in \mathcal{U}} \sum\limits_{j \in \mathcal{S}}    x_{ij} \log \left( \varepsilon R_{ij}\right)  + \sum\limits_{j \in \mathcal{T}} x_{ij} \log \left( (1 - \varepsilon) R_{ij}\right)   }{}{}\label{OPT_PB_1}
\addConstraint{x_{ij}}{\in \{0,1\}, \; i \in \mathcal{U}, j \in \mathcal{B}}{}\label{PB1_const1}
\addConstraint{\sum \limits_{j} x_{ij}}{= 1, \; \forall i \in \mathcal{U}}{}\label{PB1_const2}
\addConstraint{\sum \limits_{i} x_{ij}}{ = k_j, \; \forall j \in \mathcal{B}}{}\label{PB1_const3}
\addConstraint{\sum \limits_{j} k_{j}}{= K}{}\label{PB1_const4}
\addConstraint{p_j}{\leq p_{j}^{\rm MAX}, \; \forall j \in \mathcal{B}}{
}\label{PB1_const5}
\addConstraint{\sum \limits_{j} x_{ij} p_j  \beta_{ij} }{\geq p_{\rm min}, \; \forall i \in \mathcal{U}}{
}\label{PB1_const6}
\addConstraint{\varepsilon}{ \in \left[ 0,1 \right], }{
}\label{PB1_const7}
 \end{maxi!}
\noindent where  
p $ = \left[ p_1, \dots, p_{M+N} \right]^T$ is the vector representing the transmit power at each BS, 
k $ = \left[ k_1, \dots, k_{M+N}\right]^T$ is the vector which shows the number of UEs associated to each BS, 
and X $= \left[x_{ij}\right]_{i\in \mathcal{U},j \in \mathcal{B}}$ is the binary association matrix.\\
The artificial inclusion of vector $k$ will later allow us to determine whether a BS is overloaded or not.
Constraint \eqref{PB1_const2} ensures that each UE is associated with a unique BS, 
while constraint \eqref{PB1_const4} indicates that all UEs in the network must be served.
Furthermore, the maximum transmit power allocated per RE in each BS $j$ is restricted by $p_j^{\rm MAX}$ in constraint \eqref{PB1_const5}. 
Finally, constraint \eqref{PB1_const6} ensures the coverage of the entire network by imposing that the minimum RSRP for each UE is greater than a set threshold $p_{\rm min}$.

\section{Utility optimization using Lagrange multipliers}
\label{sec:UtilityOptimization}

In this section, 
we study the solution to our optimization problem \eqref{OPT_PB_1}.
Due to the nature of X, 
this is a mixed discrete optimization problem, 
hence complex to solve. 
To simplify the problem, 
we will first optimize the UE-BS association and the bandwidth allocation considering fixed transmit power, 
similarly to \cite{Shen_2014}. 
Then, we will optimize the transmit power level considering the first two parameters fixed. 

\subsection{Utility optimization under fixed transmit power}

Since the transmit power is fixed, 
we consider Problem \eqref{OPT_PB_1} without the Constraint \eqref{PB1_const5}.
We can solve this problem using the Lagrange multipliers, 
as it has been proposed in \cite{Ye_2013,Shen_2014}.
We introduce $\lambda = \left[ \lambda_{1}, \dots , \lambda_{K} \right]^T$, $\mu = \left[ \mu_{1}, \dots, \mu_{M+N} \right]^T$, $\alpha$, and $\rho$ as the dual variables for constraints \eqref{PB1_const6},\eqref{PB1_const3},\eqref{PB1_const4} and \eqref{PB1_const7}, respectively.
The Lagrangian function is then:

\begin{equation}\label{Lagrange_formula}
\begin{split}
    &\mathcal{L}\left( X,k,\varepsilon,\lambda,\mu,\alpha,\rho \right) = \rho \left( 1 - \varepsilon \right) - \alpha \left( \sum\limits_{ j \in \mathcal{B}} k_j - K \right) \\ &+ \sum\limits_{i} \biggl(  \sum\limits_{j \in \mathcal{S}} x_{ij} \log \left( \varepsilon R_{ij}\right) +  \sum\limits_{j \in \mathcal{T}} x_{ij} \log \left( (1 - \varepsilon) R_{ij}\right) \biggr) \\ &+ \sum\limits_{i} \lambda_i \left(
    \sum\limits_{j \in \mathcal{B}} x_{ij}p_j \beta_{ij}  - p_{\rm min}\right) + \sum\limits_{j \in \mathcal{B}} \mu_j \left( k_j - \sum\limits_i x_{ij} \right).
\end{split}
\end{equation}
After this, we are able to compute the derivative of the Lagrangian with respect to all the variables that we want to optimize, 
i.e. $x_{ij}, k_j$ and $\varepsilon$, as

\begin{equation}\label{Lagrange_derivee_xij}
\begin{split}
    \frac{\partial \mathcal{L}}{\partial x_{ij}} =
    \begin{cases}
      \log \left( \varepsilon R_{ij}\right) + \lambda_i p_j \beta_{ij}  - \mu_j , & \text{if}\ j \in \mathcal{S}, \\
      \log \left( (1 - \varepsilon) R_{ij}\right) + \lambda_i p_j \beta_{ij}  - \mu_j, & \text{otherwise}.
    \end{cases}
\end{split}
\end{equation}

\noindent The choice of the BS association is made by finding which one maximizes the derivative. 
Therefore, we can derive the following expression:

\begin{equation}\label{Lagrange_opt_xij}
\begin{split}
    x_{ij}^* =
    \begin{cases}
      1, & \text{if} \ j= \arg\max\limits_{j^\prime} \frac{\partial \mathcal{L}}{\partial x_{ij^{\prime}}},\\
      0, & \text{otherwise}.
    \end{cases}
\end{split}
\end{equation}

\noindent This association criterion is actually quite intuitive. 
Indeed, as we will see later, 
the dual variable $\mu_j$ represents the cost of association to BS $j$. 
Each UE is thus associated to the BS, 
which maximizes the difference between the data rate and the cost of association.
For the vector k, we derive its optimal value by computing the partial derivative of the Lagrangian, 
and finding its root:
\begin{equation}\label{Lagrange_opt_kj}
     k_{j}^* = e^{\mu_j -  \alpha - 1 }.
\end{equation}

\noindent Finally, we isolate all the terms of the Lagrangian function related to $\varepsilon$, 
and then compute the partial derivative with respect to this parameter as:
\begin{equation}\label{Lagrange_derivee_epsilon}
\begin{split}
     \frac{\partial \mathcal{L}}{\partial \varepsilon} &= \frac{\partial}{\partial \varepsilon} \left( \sum\limits_i \left(  \sum\limits_{j \in \mathcal{S}}    x_{ij}\log(\varepsilon) + \sum\limits_{j \in \mathcal{T}} x_{ij} \log(1 - \varepsilon) \right)  - \varepsilon\rho \right) \\
     &= \frac{1}{\varepsilon}\left( \sum\limits_i   \sum\limits_{j \in \mathcal{S}}    x_{ij} \right) - \frac{1}{1 - \varepsilon}\left( \sum\limits_i \sum\limits_{j \in \mathcal{T}} x_{ij} \right) - \rho \\ 
     &= \frac{1}{\varepsilon} K_{\mathcal{S}} - \frac{1}{1 - \varepsilon}\left( K - K_{\mathcal{S}} \right) - \rho,
\end{split}
\end{equation}
\noindent where $K_{\mathcal{S}}$ represents the number of UEs associated to a satellite in the network.
Equating \eqref{Lagrange_derivee_epsilon} to 0, 
we obtain:

\begin{equation}\label{Lagrange_epsilon_polynome}
\rho \varepsilon^2 - \left( K + \rho \right) \varepsilon + K_{\mathcal{S}} = 0,
\end{equation}
which allows us to find the following optimal value: 
\begin{equation}\label{Lagrange_epsilon_optimal}
\varepsilon^* = \frac{K + \rho - \sqrt{ \left(  K + \rho \right)^2 - 4 \rho K_{\mathcal{S}}}}{2\rho}.
\end{equation}

\noindent From eq. \eqref{Lagrange_epsilon_optimal},
we can observe that the proportion of bandwidth allocated to the Non-Terrestrial tier is directly proportional to the number of UEs associated with an LEO satellite.
In fact, if we gradually increase the value of $K_{\mathcal{S}}$ from $0$ to $K$, 
the value of $\varepsilon^*$ slowly shifts from $0$ to $1$. 
We thereby introduce the Lagrangian dual function, 
which can be written as:

\begin{equation}\label{Dual_def}
\begin{split}
    &\mathcal{D}\left(\lambda,\mu,\alpha,\rho \right) = \max\limits_{X,k,\varepsilon} \mathcal{L}\left(   X,k,\varepsilon,\lambda,\mu,\alpha,\rho \right).
\end{split}
\end{equation}

\noindent Accordingly, the Lagrangian problem \eqref{OPT_PB_1} can then be rewritten as:
\begin{equation}\label{Dual_equation}
\begin{split}
    & \min\limits_{\mu, \lambda, \alpha, \rho}    \mathcal{D}\left(\lambda,\mu,\alpha,\rho \right) .
\end{split}
\end{equation}

\noindent By injecting the expressions obtained in \eqref{Lagrange_opt_xij}, \eqref{Lagrange_opt_kj}, and \eqref{Lagrange_epsilon_optimal}, 
we get:

\begin{equation}\label{Dual_equation_long}
\begin{split}
    &\mathcal{D}\left(\lambda,\mu,\alpha,\rho \right) = \mathcal{L}\left(   X^*,k^*,\varepsilon^*,\lambda,\mu,\alpha,\rho \right) \\
    &= \sum\limits_{i} \biggl( \sum\limits_{j \in \mathcal{S}}    x_{ij}^* \log \left( \varepsilon^* R_{ij}\right)  +  \sum\limits_{j \in \mathcal{T}} x_{ij}^* \log \left( (1 - \varepsilon^*) R_{ij}\right) \biggl) \\ &+ \sum\limits_{i} \lambda_i \left(
    \sum\limits_j x_{ij}^*p_j \beta_{ij} - p_{\rm min}\right) + \sum\limits_j \mu_j \left( k_j^* - \sum\limits_i x_{ij}^* \right) \\ &+ \rho \left( 1 - \varepsilon^* \right) - \alpha \left( \sum\limits_j k_j^* - K \right).
\end{split}
\end{equation}

\noindent In order to minimize this function, we use the subgradient method to update the Lagrange multipliers, 
as already suggested in \cite{Ye_2013,Shen_2014}, as follows:

\begin{equation}\label{Dual_update_mu}
\mu_j\left( t + 1 \right) = \mu_j\left( t \right) - \delta_1(t) \left( k_j^* - \sum\limits_i x_{ij}^* \right),
\end{equation}

\begin{equation}\label{Dual_update_lambda}
\lambda_i\left( t + 1 \right) = \lambda_i\left( t \right) - \delta_2(t) \left(\sum\limits_j x_{ij}^*p_j \beta_{ij}  - p_{\rm min} \right),
\end{equation}

\begin{equation}\label{Dual_update_alpha}
\alpha\left( t + 1 \right) = \alpha\left( t \right) - \delta_3(t) \left(K - \sum\limits_j k_{j}^* \right),
\end{equation}

\begin{equation}\label{Dual_update_rho}
\rho\left( t + 1 \right) = \rho\left( t \right) + \delta_4(t) \varepsilon^*,
\end{equation}
where $\delta_1(t)$, $\delta_2(t)$, $\delta_3(t)$, and $\delta_4(t)$ represent the step-sizes used for each dual variable.
Since the dual problem is always convex, 
the usage of the subgradient method with decreasing step sizes guarantees convergence to the optimal solution of this problem \cite{boyd_subgradient}.
\newline
Eq. \ref{Dual_update_mu} explains how the proposed framework balances the load among the BSs. Indeed, as stated previously, 
$\mu_j$ is the cost of association to BS $j$. 
This price will only rise if the right component in the equation is negative, 
meaning that the number of UEs associated to the BS is excessively large. 
This way, a BS with fewer UEs has a lower cost and it is more attractive, and vice-versa.

\subsection{Transmit power optimization under fixed association}

Once the UE-BS association and bandwidth allocation problem has been solved, 
we fix X and $\varepsilon$ to further optimize the transmit power at each BS and maximize the log-throughput of the network.
For ease of reading, 
we will denote by $f\left( p \right)$ the sum log-throughput of the network \eqref{OPT_PB_1} to indicate that it is a function of the transmit power vector p.
The transmit power optimization problem can then be expressed as:

\begin{maxi!}|s|[2]
{p}{f\left(p\right)}{}{}\label{OPT_PB_2}
\addConstraint{\sum \limits_{j} x_{ij} p_j  \beta_{ij} }{\geq p_{\rm min}, \; \forall i \in \mathcal{U}}{
}\label{PB2_const1}
\addConstraint{p_j}{\leq p_{j}^{MAX}, \; \forall j \in \mathcal{B}.}{
}\label{PB2_const2}
 \end{maxi!}
 
\noindent Since the objective function is concave w.r.t. p, 
we can try to approximate the zero of the gradient using the Newton-Raphson iterative method to maximize the utility function, 
as demonstrated in \cite{Shen_2014}.
As indicated in \cite{WeiYu2013}, 
it is also possible to use only diagonal entries of the Hessian matrix to reduce the computational complexity of inverting it.
To this end, 
the first and second order derivatives are computed as follows:

\begin{equation}
\begin{split}
\frac{\partial f(p)}{\partial p_j} &=\sum_i \frac{\gamma_{i j}}{r_{i j}\left(1 + \gamma_{i j}\right)} \frac{x_{i j}}{p_j} \\
&- \sum_i \sum_{j^{\prime} \neq j} \frac{\beta_{i j} \gamma_{i j^{\prime}}^2}{ \beta_{i j^{\prime}}  r_{i j^{\prime}}\left(1 + \gamma_{i j^{\prime}}\right)} \frac{x_{i j^{\prime}}}{p_{j^{\prime}}},
\end{split}
\end{equation}
and
\begin{equation}
\begin{split}
\frac{\partial^2 f(p)}{\partial p_j^2}&=-\sum_i\left(\frac{1}{r_{i j}^2}+\frac{1}{r_{i j}}\right) \frac{\gamma_{i j}^2}{\left(1 + \gamma_{i j}\right)^2} \frac{x_{i j}}{p_j^2} \\
&+\sum_i \sum_{j^{\prime} \neq j} \frac{\beta_{i j}^2 \gamma_{i j^{\prime}}^3\left(2 r_{i j^{\prime}}+ \gamma_{i j^{\prime}}\left(r_{i j^{\prime}}-1\right)\right)}{\beta_{i j^{\prime}}^2 r_{i j^{\prime}}^2\left(1 + \gamma_{i j^{\prime}}\right)^2} \frac{x_{i j^{\prime}}}{p_{j^{\prime}}^2},
\end{split}
\end{equation}
where
\begin{equation}
r_{ij} = \log\left( 1 + \gamma_{ij} \right).
\end{equation}
The Newton step is then:
\begin{equation}\label{eq_delta_newton}
\Delta p_j = \frac{\partial f(p)}{\partial p_j} \Bigg/ \left\vert \frac{\partial^2 f(p)}{\partial p_j^2} \right\vert.
\end{equation}
Once we update the transmit power vector using \eqref{eq_delta_newton}, it is necessary to project the value in a region where constraints \eqref{PB2_const1} and \eqref{PB2_const2} are respected.
Naturally, the upper bound of our feasible region is the maximum transmit power for each BS.
For the lower bound, 
we utilize the minimal coverage constraint, 
i.e. we know that for a BS $j$, 
all UEs associated to it should be receiving a signal power greater than $p_{\rm min}$.
This can be translated as:
\begin{equation}
\begin{split}
& \forall i \in \mathcal{U}_j,\quad p_j \quad {\geq} \quad \frac{p_{\rm min}}{ \beta_{ij}},
\end{split}
\end{equation}
with $\mathcal{U}_j$ being the set of UEs associated to the BS $j$.
We are therefore able to establish the lower bound of the feasibility region for each BS $j$ as:
\begin{equation}\label{tau_j}
\tau_j=\max _{i \in \mathcal{U}_j}\left(\frac{p_{\min }}{\beta_{ij}}\right).
\end{equation}

\noindent Finally, the transmit power update done at the end of step $t$ is written as such:
\begin{equation}\label{power_update}
p_j^{(t+1)}=\left[p_j^{(t)}+\delta_5(t) \Delta p_j\right]_{\tau_j}^{p_{j}^{\rm MAX}},
\end{equation}
with $\delta_5(t)$ being a step-size factor.


\section{Simulation results and analysis}
\label{sec:Simulation_Results}
In this section, 
we assess the effectiveness of our proposed optimization framework for UE association, bandwidth allocation, and transmit power control in an integrated TN-NTN.
We analyse a rural scenario where the macro BSs are deployed in an hexagonal grid layout \cite{3GPPTR36.942}.
{Without loss of generality, we consider an LEO constellation employing earth-fixed beams, such that, at a given instant, a UE can only be served by a unique satellite, and we restrict our study to an area of $2 500$ $\mathrm{km}^2$, corresponding to the coverage provided by the beam of an LEO satellite \cite{3GPPTR38.821}.}

\begin{table}[h!]
\begin{center}
\begin{tabular}{|l|l|}
\hline Parameter & Value \\
\hline Total Bandwidth $W$  & $40$ $\mathrm{MHz}$ \\
\hline Carrier frequency $f_c$  & $2$ $\mathrm{GHz}$ \\
\hline Subcarrier Spacing   & $15$ $\mathrm{kHz}$ \\
\hline UE density & $2 \text{ UE} / \text{km}^2$ \\
\hline Inter-Site Distance & $1732$ $m$ \\
\hline Number of Macro BSs & $1067$ \\
\hline Terrestrial Max Tx Power per RE $p_{j}^{MAX}$ \cite{3GPPTR36.814} & $ 17.7 \text{ } \mathrm{dBm}$ \\
\hline Satellite Max Tx Power per RE $p_{j}^{MAX}$ \cite{3GPPTR38.821} & $15.8 \text{ } \mathrm{dBm}$ \\
\hline Antenna gain (Terrestrial) $G_{T_X}$ \cite{3GPPTR36.931} & $14\text{ } \mathrm{dBi}$ \\
\hline Antenna gain (Satellite) $G_{T_X}$ \cite{3GPPTR38.821} & $30\text{ } \mathrm{dBi}$ \\
\hline Shadowing Loss (Terrestrial) $SF$ \cite{3GPPTR38.901} & $4\text{ } - \text{ } 8\text{ } \mathrm{dB}$ \\
\hline Shadowing Loss (Satellite) $SF$ \cite{3GPPTR38.811} & $0\text{ } - \text{ } 12\text{ } \mathrm{dB}$ \\
\hline Line-of-Sight Probability (Satellite / Terrestrial) & Provided in \cite{3GPPTR38.811} / \cite{3GPPTR38.901} \\
\hline White Noise Power Density \cite{3GPPTR36.931} & $-174 \text{ }$ $ \mathrm{dBm} /  \mathrm{Hz}$ \\
\hline Coverage threshold $p_{\rm min}$ & $-120 \text{ }$ $ \mathrm{dBm}$ \\
\hline
\end{tabular}
\end{center}
\caption{Simulation parameters.}
\label{simul_params}
\end{table}


\noindent As for the UE deployment, 
we consider an inhomogeneous deployment.  
Indeed, we first randomly select $30\%$ of the macro BSs,
and for these macro BSs, 
we deploy the UEs in a "hot-spot" manner, 
to possibly create overload in the related cells and allow our framework to demonstrate its effectiveness.
Half of the UEs are deployed among those hot-spots, 
and the other half are uniformly spread across the entire area.
The most important simulation parameters, set according to  \cite{3GPPTR36.763, 3GPPTR38.811, 3GPPTR38.821,3GPPTR38.901, 3GPPTR36.814, 3GPPTR36.931}, are listed in Table 
\ref{simul_params}.
We compare the performance of our framework with two different benchmarks:
The \texttt{Baseline} relates to a standard terrestrial network where a bandwidth of $10$ MHz is available at the macro BSs \cite{3GPPTR36.814} and the \texttt{3GPP} NTN scenario where the bandwidth $W$ is split accordingly to the 3GPP recommendations \cite{3GPPTR38.821}, i.e., $30$ MHz allocated to the satellite and $10$ MHz allocated to the macro BSs.
Note that in both benchmarks the UEs associate to the BS providing the largest RSRP.

\subsection{Framework convergence analysis} \label{SubSeq : Framework Convergence Analysis}

\begin{figure}[h]
    \centering
    \includegraphics[width= 0.5\textwidth ]{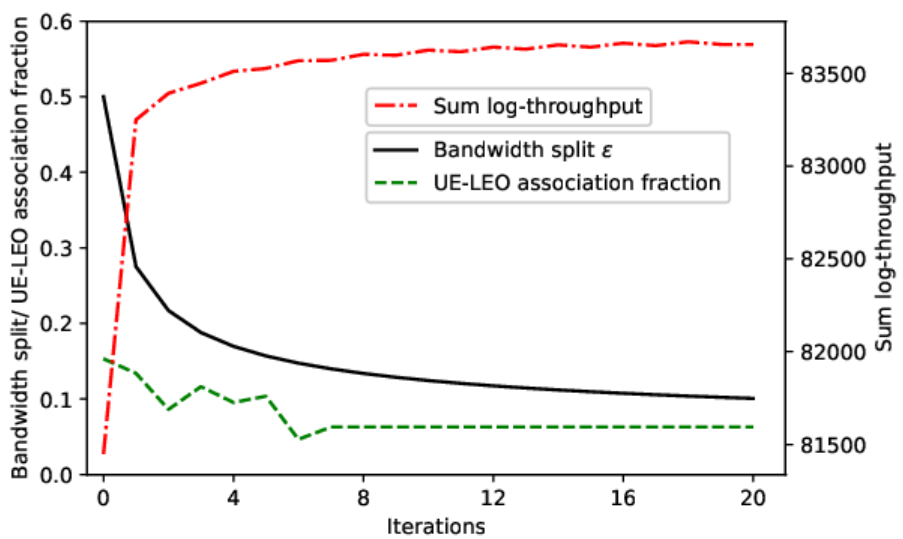}
    \caption{Evolution of the bandwidth allocation proportion and the utility function of our framework.} 
    \label{triple_plot}
    \vspace{-0.2cm}
\end{figure}

In this section, 
we analyse the convergence of the proposed optimization framework. 
Specifically, 
Figure \ref{triple_plot} shows the iterative evolution of 
1) the network SLT, 
along with 
2) the optimal bandwidth split ($\varepsilon$) and 
3) the actual fraction of UEs associated to the satellite in the network, 
i.e. $\frac{k_0}{\sum\limits_{j} k_j}$.

\noindent When initializing the algorithm presented in Sec. \ref{sec:UtilityOptimization}, 
the bandwidth split, $\varepsilon$ is set to $0.5$,
and the fraction of UEs associated to the LEO satellite is $0.15$. The initial UE-BS association follows the max-RSRP rule.
During the iterative process, 
we can see that the log-throughput continuously improves, 
while the ratio of UEs associated to the LEO satellite and the fraction of bandwidth allocated to it decrease. 
Eventually, the algorithm converges after $20$ iterations, 
with an improved SLT 
and $11 \%$ of the bandwidth being allocated to the LEO satellite and approximately $6 \%$ of the UEs associated to it. 
Note that this bandwidth split is different than the one recommended in \texttt{3GPP} specifications \cite{3GPPTR38.821}.
For readability purposes, we chose not to display the evolution of the transmit power. Considering that all BSs were initially transmitting at their maximum power, we observe an $82\%$ decrease of the average transmit power. This is explained by the fact that our framework reduces the transmit power of the BSs that have no UEs and thus no coverage constraint \eqref{PB1_const6} to uphold.
In the following, 
we denote by $\varepsilon_{opt}$ the optimal bandwidth split derived by the proposed framework.

\subsection{Network coverage analysis}

\begin{figure}[h]
    \centering
    \includegraphics[width=0.45\textwidth]{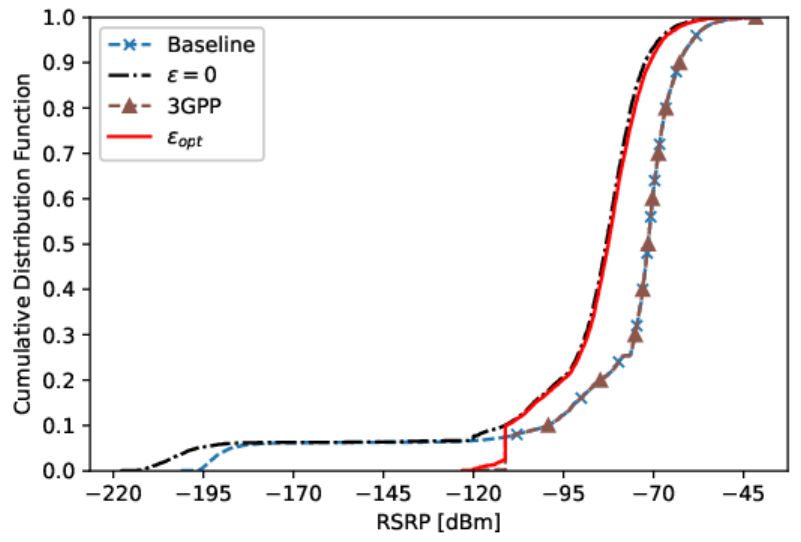}
    \caption{CDFs of the UE RSRP, with ($\varepsilon_{opt}$, \texttt{3GPP}) and without ($\varepsilon = 0$, \texttt{Baseline}) an active satellite.} 
    \label{cdf_rsrp}
\end{figure}

In this section, 
we study the benefits of integrating an NTN to a terrestrial network in terms of the coverage,
i.e. the capability to provide wireless services.
Figure \ref{cdf_rsrp} shows the cumulative distribution function (CDF) of the RSRP perceived at each UE from the serving BS in four different scenarios: 1) the \texttt{Baseline} setting, 2) a scenario where all the bandwidth is allocated to the terrestrial network ($\varepsilon = 0$) and the UE association and power control is done through our framework, 3) the \texttt{3GPP} setting and 4) the scenario where bandwidth split, association, and power are allocated through the proposed framework ($\varepsilon_{opt}$).
In the first two scenarios, we observe a similar performance in terms of coverage since all the UEs are served by macro BSs, which leads to $7 \%$ of the UEs to be out of coverage since their respective RSRPs are below the threshold $p_{\rm min}$.
In contrast, when integrating the NTN in the last two scenarios, 
the proportion of UEs out of coverage drastically drops down to around $0.4$ $\%$.
Indeed, the satellite can reach UEs located at the cell edge and provide them with a signal of much better quality than that provided by the strongest macro BSs.


\subsection{UE rate analysis}

\begin{figure}[t]
    \centering
\includegraphics[width=0.5\textwidth]{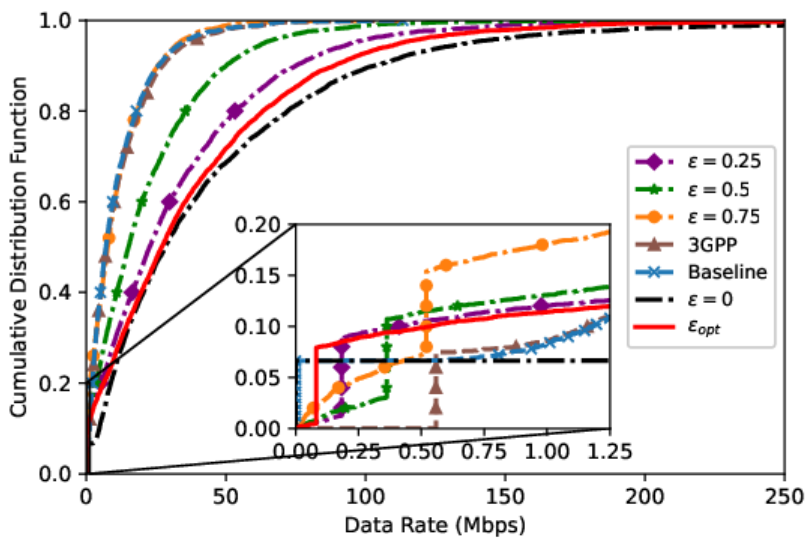}
    \caption{CDFs of UE data rates for various bandwidth allocation settings.} 
    \label{throughput_cdfs}
\end{figure}

In this section, we compare the data rate performance achieved by our framework with the one of the \texttt{Baseline} and \texttt{3GPP} settings. Also, we consider the case where only the user association and the power control are optimized and the bandwidth split is fixed, i.e., $\varepsilon \in \{0, 0.25, 0.5, 0.75\}$.
Figure \ref{throughput_cdfs} shows the CDFs of the data rate achieved when considering the various deployment and resource allocation scenarios.
Also, Table \ref{Table_perf} presents the $5$-th percentile, the mean, the median, and the $95$-th percentile of the different data rate distributions resulting from the most relevant of the compared solutions. 
\begin{table}[h]
\begin{center}
\begin{tabular}{|l|l|l|l|l|l|}
\hline  & $\varepsilon = 0$ & $\varepsilon_{opt}$&$\varepsilon$ = 0.75 & 3GPP & Baseline\\ 
\hline $5$-th \%ile (kbps) & 0  & $81$ & $614$ & $558$ & $0$  \\ 
\hline Mean (Mbps) & $44.4$ & $38.3$ & $12.0$ & $11.7$ & $11.1$  \\ 
\hline Median (Mbps) & $28.3$ & $27.1$ & $7.7$ & $7.3$ & $7.1$ \\ 
\hline $95$-th \%tile (Mbps) & $136.6$  & $112.7$ & $37$ & $36.4$ & $34.1$  \\
\hline
\end{tabular}
\end{center}
\caption{Data-rate analysis.}
\label{Table_perf}
\end{table}

\noindent We first notice that higher data rates, in average, are achieved when we allocate a large split of the bandwidth to the terrestrial network. 
This is because of the large spectrum reuse in the area under investigation. 
However, it is important to note that the tail of the rate distribution greatly suffers if we prioritize the terrestrial network when controlling the spectrum split.
With the \texttt{Baseline} setting and when all the bandwidth is allocated to macro BSs ($\varepsilon = 0$), 
around $7 \%$ of the UEs are out of coverage,  as we observed in the previous section, and their rate is null. 
When the NTN bandwidth is increased, 
the coverage holes of the network are reduced, 
and the rate experienced by the cell edge UEs 
increases. This can be observed in the zoom of Fig. \ref{throughput_cdfs}.
Overall, 
we can highlight the underlying trade-off between cell-edge (5-th percentile UEs) and cell-center (mean/median and 95-th percentile UEs) throughput.
If the operator gives a small share of the bandwidth to the satellite, 
it may achieve large cell-center UE data rates at the expense of coverage holes and degraded performance at the cell edge.
In contrast, 
if the operator decides to allocate a large share of the bandwidth to the satellite, 
the cell edge performance greatly improves, 
at the expense of the cell-center UE data rate. 
For example, 
a UE which would be out of coverage when $\varepsilon = 0$ or in the \texttt{Baseline} scenario experiences a data rate of roughly $81 \text{ } \mathrm{kbps}$ if, using the proposed framework, we optimally set $\varepsilon$ to $0.11$.
With our proposal, the mean UE rate decreases by $14\%$ with respect to the setting of $\varepsilon = 0$ but results in a gain of more than 200 $\%$ with respect to the \texttt{Baseline} and the \texttt{3GPP} settings. 
Therefore, our framework is able to find the best solution to this trade-off by improving the coverage condition of the UEs that suffer from large path losses whilst providing large data rates to cell-center UEs.

\section{Conclusion}
\label{sec:conclusions}

In this paper, we have studied the throughput-coverage trade-off in a hybrid network comprised of terrestrial macro BSs and an LEO satellite. We have proposed a framework to control the UE association, transmission power, and bandwidth allocation between terrestrial and satellite BSs. 
Our proposal is able to distribute the load, 
while mitigating the number of coverage holes and maximizing the SLT in the network.
Specifically, we 
demonstrated that by incorporating an LEO satellite on top of the terrestrial network in rural areas,
the proportion of UEs out-of-coverage significantly drops down. Also, by studying the scenario where both tiers share the bandwidth,
we were able to underline the trade-off between minimizing the coverage holes and enhancing the maximum throughput of the network,
and strike the optimal point.
Finally, our results indicate that the UE-BS association resulting from our framework greatly improves the performance of the network in terms of mean and 95-th percentile of throughput compared to the max-RSRP rule.
Our analysis highlights the critical role that NTNs will play in the following years in providing reliable service throughout the world.
Our future works will include an analysis from an energy efficiency point of view, 
as well as a refinement of the framework presented in this paper.

\bibliographystyle{IEEEtran}
\bibliography{main.bib}

\begin{acronym}[AAAAAAAAA]

 \acro{2D}{two-dimensional}
 \acro{3D}{three-dimensional}
 \acro{3G}{third generation}
 \acro{3GPP}{third generation partnership project}
 \acro{4G}{fourth generation}
 \acro{5G}{fifth generation}
 \acro{5GC}{5G Core Network}
 \acro{AAA}{authentication, authorisation and accounting}
 \acro{ABS}{almost blank subframe}
 \acro{AC}{alternating current}
 \acro{ACIR}{adjacent channel interference rejection ratio}
 \acro{ACK}{acknowledgment}
 \acro{ACL}{allowed CSG list}
 \acro{ACLR}{adjacent channel leakage ratio}
 \acro{ACPR}{adjacent channel power ratio}
 \acro{ACS}{adjacent channel selectivity}
 \acro{ADC}{analog-to-digital converter}
 \acro{ADSL}{asymmetric digital subscriber line}
 \acro{AEE}{area energy efficiency}
 \acro{AF}{amplify-and-forward}
 \acro{AGCH}{access grant channel}
 \acro{AGG}{aggressor cell}
 \acro{AH}{authentication header}
 \acro{AI}{artificial intelligence}
 \acro{AKA}{authentication and key agreement}
 \acro{AMC}{adaptive modulation and coding}
 \acro{ANN}{artificial neural network}
 \acro{ANR}{automatic neighbour relation}
 \acro{AoA}{angle of arrival}
 \acro{AoD}{angle of departure}
 \acro{APC}{area power consumption}
 \acro{API}{application programming interface}
 \acro{APP}{a posteriori probability}
 \acro{AR}{augmented reality}
 \acro{ARIMA}{autoregressive integrated moving average}
 \acro{ARQ}{automatic repeat request}
 \acro{AS}{access stratum}
 \acro{ASE}{area spectral efficiency}
 \acro{ASM}{advanced sleep mode}
 \acro{ASN}{access service network}
 \acro{ATM}{asynchronous transfer mode}
 \acro{ATSC}{Advanced Television Systems Committee}
 \acro{AUC}{authentication centre}
 \acro{AWGN}{additive white gaussian noise}
 \acro{BB}{baseband}
 \acro{BBU}{baseband unit}
 \acro{BCCH}{broadcast control channel}
 \acro{BCH}{broadcast channel}
 \acro{BCJR}{Bahl-Cocke-Jelinek-Raviv} 
 \acro{BE}{best effort}
 \acro{BER}{bit error rate}
 \acro{BLER}{block error rate}
 \acro{BPSK}{binary phase-shift keying}
 \acro{BR}{bit rate}
 \acro{BS}{base station}
 \acro{BSC}{base station controller}
 \acro{BSIC}{base station identity code}
 \acro{BSP}{binary space partitioning}
 \acro{BSS}{blind source separation}
 \acro{BTS}{base transceiver station}
 \acro{BWP}{Bandwidth Part}
 \acro{CA}{carrier aggregation}
 \acro{CAC}{call admission control}
 \acro{CaCo}{carrier component}
 \acro{CAPEX}{capital expenditure}
 \acro{capex}{capital expenses}
 \acro{CAS}{cluster angular spread}
 \acro{CATV}{community antenna television}
 \acro{CAZAC}{constant amplitude zero auto-correlation}
 \acro{CC}{component carrier}
 \acro{CCCH}{common control channel}
 \acro{CCDF}{complementary cumulative distribution function}
 \acro{CCE}{control channel element}
 \acro{CCO}{coverage and capacity optimisation}
 \acro{CCPCH}{common control physical channel}
 \acro{CCRS}{coordinated and cooperative relay system}
 \acro{CCTrCH}{coded composite transport channel}
 \acro{CDF}{cumulative distribution function}
 \acro{CDMA}{code division multiple access}
 \acro{CDS}{cluster delay spread}
 \acro{CESM}{capacity effective SINR mapping}
 \acro{CO$_{2e}$}{carbon dioxide equivalent}
 \acro{CFI}{control format indicator}
 \acro{CFL}{Courant-Friedrichs-Lewy}
 \acro{CGI}{cell global identity}
 \acro{CID}{connection identifier}
 \acro{CIF}{carrier indicator field}
 \acro{CIO}{cell individual offset}
 \acro{CIR}{channel impulse response}
 \acro{CNN}{Convolutional Neural Network}
 \acro{CMF}{cumulative mass function}
 \acro{C-MIMO}{cooperative MIMO}
 \acro{CN}{core network}
 \acro{COC}{cell outage compensation}
 \acro{COD}{cell outage detection}
 \acro{CoMP}{coordinated multi-point}
 \acro{ConvLSTM}{Convolutional LSTM}
 \acro{CP}{cycle prefix}
 \acro{CPC}{cognitive pilot channel}
 \acro{CPCH}{common packet channel}
 \acro{CPE}{customer premises equipment}
 \acro{CPICH}{common pilot channel}
 \acro{CPRI}{common public radio interface}
 \acro{CPU}{central processing unit}
 \acro{CQI}{channel quality indicator}
 \acro{CR}{cognitive radio}
 \acro{CRAN}{centralized radio access network} 
 \acro{C-RAN}{cloud radio access network} 
 \acro{CRC}{cyclic redundancy check}
 \acro{CRE}{cell range expansion}
 \acro{C-RNTI}{cell radio network temporary identifier}
 \acro{CRP}{cell re-selection parameter}
 \acro{CRS}{cell-specific reference symbol}
 \acro{CRT}{cell re-selection threshold}
 \acro{CSCC}{common spectrum coordination channel}
 \acro{CSG ID}{closed subscriber group ID}
 \acro{CSG}{closed subscriber group}
 \acro{CSI}{channel state information}
 \acro{CSIR}{receiver-side channel state information}
 \acro{CSI-RS}{channel state information-reference signals}
 \acro{CSO}{cell selection offset}
 \acro{CTCH}{common traffic channel}
 \acro{CTS}{clear-to-send} 
 \acro{CU}{central unit}
 \acro{CV}{cross-validation}
 \acro{CWiND}{Centre for Wireless Network Design}
 \acro{D2D}{device to device}
 \acro{DAB}{digital audio broadcasting}
 \acro{DAC}{digital-to-analog converter}
 \acro{DAS}{distributed antenna system}
 \acro{dB}{decibel}
 \acro{dBi}{isotropic-decibel}
 \acro{DC}{direct current}
 \acro{DCCH}{dedicated control channel}
 \acro{DCF}{decode-and-forward}
 \acro{DCH}{dedicated channel}
 \acro{DC-HSPA}{dual-carrier high speed packet access}
 \acro{DCI}{downlink control information}
 \acro{DCM}{directional channel model}
 \acro{DCP}{dirty-paper coding}
 \acro{DCS}{digital communication system}
 \acro{DECT}{digital enhanced cordless telecommunication}
 \acro{DeNB}{donor eNodeB}
 \acro{DFP}{dynamic frequency planning}
 \acro{DFS}{dynamic frequency selection}
 \acro{DFT}{discrete Fourier transform}
 \acro{DFTS}{discrete Fourier transform spread}
 \acro{DHCP}{dynamic host control protocol}
 \acro{DL}{downlink}
 \acro{DMC}{dense multi-path components}
 \acro{DMF}{demodulate-and-forward}
 \acro{DMT}{diversity and multiplexing tradeoff}
  \acro{DNN}{deep neural network} 
 \acro{DoA}{direction-of-arrival}
 \acro{DoD}{direction-of-departure}
 \acro{DoS}{denial of service}
 \acro{DPCCH}{dedicated physical control channel}
 \acro{DPDCH}{dedicated physical data channel}
 \acro{D-QDCR}{distributed QoS-based dynamic channel reservation}
 \acro{DQL}{deep Q-learning}
  \acro{DRAN}{distributed radio access network}
 \acro{DRS}{discovery reference signal}
 \acro{DRL}{deep reinforcement learning}
 \acro{DRX}{discontinuous reception}
 \acro{DS}{down stream}
 \acro{DSA}{dynamic spectrum access}
 \acro{DSCH}{downlink shared channel}
 \acro{DSL}{digital subscriber line}
 \acro{DSLAM}{digital subscriber line access multiplexer}
 \acro{DSP}{digital signal processor}
 \acro{DT}{decision tree}
 \acro{DTCH}{dedicated traffic channel}
 \acro{DTX}{discontinuous transmission}
   \acro{DU}{distributed unit}
 \acro{DVB}{digital video broadcasting}
 \acro{DXF}{drawing interchange format}
 \acro{E2E}{end-to-end}
 \acro{EAGCH}{enhanced uplink absolute grant channel}
 \acro{EA}{evolutionary algorithm}
 \acro{EAP}{extensible authentication protocol}
 \acro{EC}{evolutionary computing}
 \acro{ECGI}{evolved cell global identifier}
 \acro{ECR}{energy consumption ratio}
 \acro{ECRM}{effective code rate map}
 \acro{EDCH}{enhanced dedicated channel}
 \acro{EE}{energy efficiency}
 \acro{EESM}{exponential effective SINR mapping}
 \acro{EF}{estimate-and-forward}
 \acro{EGC}{equal gain combining}
 \acro{EHICH}{EDCH HARQ indicator channel}
 \acro{eICIC}{enhanced intercell interference coordination}
 \acro{EIR}{equipment identity register}
 \acro{EIRP}{effective isotropic radiated power}
 \acro{ELF}{evolutionary learning of fuzzy rules}
 \acro{eMBB}{enhanced mobile broadband}
  \acro{EMR}{Electromagnetic-Radiation}
 \acro{EMS}{enhanced messaging service}
 \acro{eNB}{evolved NodeB}
 \acro{eNodeB}{evolved NodeB}
 \acro{EoA}{elevation of arrival}
 \acro{EoD}{elevation of departure}
 \acro{EPB}{equal path-loss boundary}
 \acro{EPC}{evolved packet core}
 \acro{EPDCCH}{enhanced physical downlink control channel}
 \acro{EPLMN}{equivalent PLMN}
 \acro{EPS}{evolved packet system}
 \acro{ERAB}{eUTRAN radio access bearer}
 \acro{ERGC}{enhanced uplink relative grant channel}
 \acro{ERTPS}{extended real time polling service}
 \acro{ESB}{equal downlink receive signal strength boundary}
 \acro{ESF}{even subframe}
 \acro{ESP}{encapsulating security payload}
 \acro{ETSI}{European Telecommunications Standards Institute}
 \acro{E-UTRA}{evolved UTRA}
 \acro{EU}{European Union}
 \acro{EUTRAN}{evolved UTRAN}
 \acro{EVDO}{evolution-data optimised}
 \acro{FACCH}{fast associated control channel}
 \acro{FACH}{forward access channel}
 \acro{FAP}{femtocell access point}
 \acro{FARL}{fuzzy assisted reinforcement learning}
 \acro{FCC}{Federal Communications Commission}
 \acro{FCCH}{frequency-correlation channel}
 \acro{FCFS}{first-come first-served}
 \acro{FCH}{frame control header}
 \acro{FCI}{failure cell ID}
 \acro{FD}{frequency-domain}
 \acro{FDD}{frequency division duplexing}
 \acro{FDM}{frequency division multiplexing}
 \acro{FDMA}{frequency division multiple access}
 \acro{FDTD}{finite-difference time-domain}
 \acro{FE}{front-end}
 \acro{FeMBMS}{further evolved multimedia broadcast multicast service}
 \acro{FER}{frame error rate}
 \acro{FFR}{fractional frequency reuse}
 \acro{FFRS}{fractional frequency reuse scheme}
 \acro{FFT}{fast Fourier transform}
 \acro{FFU}{flexible frequency usage}
 \acro{FGW}{femtocell gateway}
 \acro{FIFO}{first-in first-out}
 \acro{FIS}{fuzzy inference system}
 \acro{FMC}{fixed mobile convergence}
 \acro{FPC}{fractional power control}
 \acro{FPGA}{field-programmable gate array}
 \acro{FRS}{frequency reuse scheme}
 \acro{FTP}{file transfer protocol}
 \acro{FTTx}{fiber to the x}
 \acro{FUSC}{full usage of subchannels}
 \acro{GA}{genetic algorithm}
 \acro{GAN} {generic access network}
 \acro{GANC}{generic access network controller}
 \acro{GBR}{guaranteed bitrate}
 \acro{GCI}{global cell identity}
 \acro{STGCN}{Spatio-Temporal Graph convolutional network}
 \acro{GERAN}{GSM edge radio access network}
 \acro{GGSN}{gateway GPRS support node}
 \acro{GHG}{greenhouse gas}
 \acro{GMSC}{gateway mobile switching centre}
 \acro{gNB}{next generation NodeB}
 \acro{GNN}{Graph Neural Network}
 \acro{GNSS}{global navigation satellite system}
 \acro{GP}{genetic programming}
 \acro{GPON}{Gigabit passive optical network}
 \acro{GPP}{general purpose processor}
 \acro{GPRS}{general packet radio service}
 \acro{GPS}{global positioning system}
 \acro{GPU}{graphics processing unit}
 \acro{GRU}{gated recurrent unit}
 \acro{GSCM}{geometry-based stochastic channel models}
 \acro{GSM}{global system for mobile communication}
 \acro{GTD}{geometry theory of diffraction}
 \acro{GTP}{GPRS tunnel protocol}
 \acro{GTP-U}{GPRS tunnel protocol - user plane}
 \acro{HA}{historical average}
 \acro{HARQ}{hybrid automatic repeat request}
 \acro{HBS}{home base station}
 \acro{HCN}{heterogeneous cellular network}
 \acro{HCS}{hierarchical cell structure}
  \acro{HD}{high definition}
 \acro{HDFP}{horizontal dynamic frequency planning}
 \acro{HeNB}{home eNodeB}
 \acro{HeNodeB}{home eNodeB}
 \acro{HetNet}{heterogeneous network}
 \acro{HiFi}{high fidelity}
 \acro{HII}{high interference indicator}
 \acro{HLR}{home location register}
 \acro{HNB}{home NodeB}
 \acro{HNBAP}{home NodeB application protocol}
 \acro{HNBGW}{home NodeB gateway}
 \acro{HNodeB}{home NodeB}
 \acro{HO}{handover}
 \acro{HOF}{handover failure}
 \acro{HOM}{handover hysteresis margin}
 \acro{HPBW}{half power beam width}
 \acro{HPLMN}{home PLMN}
 \acro{HPPP}{homogeneous Poison point process}
 \acro{HRD}{horizontal reflection diffraction}
 \acro{HSB}{hot spot boundary}
 \acro{HSDPA}{high speed downlink packet access}
 \acro{HSDSCH}{high-speed DSCH}
 \acro{HSPA}{high speed packet access}
 \acro{HSS}{home subscriber server}
 \acro{HSUPA}{high speed uplink packet access}
 \acro{HUA}{home user agent}
 \acro{HUE}{home user equipment}
 \acro{HVAC}{heating, ventilating, and air conditioning}
 \acro{HW}{Holt-Winters}
 \acro{IC}{interference cancellation}
 \acro{ICI}{inter-carrier interference}
 \acro{ICIC}{intercell interference coordination}
 \acro{ICNIRP}{International Commission on Non-Ionising Radiation Protection}
 \acro{ICS}{IMS centralised service}
 \acro{ICT}{information and communication technology}
 \acro{ID}{identifier}
 \acro{IDFT}{inverse discrete Fourier transform}
 \acro{IE}{information element}
 \acro{IEEE}{Institute of Electrical and Electronics Engineers}
 \acro{IETF}{Internet engineering task force}
 \acro{IFA}{Inverted-F-antennas}
 \acro{IFFT}{inverse fast Fourier transform}
 \acro{i.i.d.}{independent and identical distributed}
 \acro{IIR}{infinite impulse response}
 \acro{IKE}{Internet key exchange}
 \acro{IKEv2}{Internet key exchange version 2}
 \acro{ILP}{integer linear programming}
 \acro{IMEI}{international mobile equipment identity}
 \acro{IMS}{IP multimedia subsystem}
 \acro{IMSI}{international mobile subscriber identity}
 \acro{IMT}{international mobile telecommunications}
 \acro{INH}{indoor hotspot}
 \acro{IOI}{interference overload indicator}
 \acro{IoT}{Internet of things}
 \acro{IP}{Internet protocol}
 \acro{IPSEC}{Internet protocol security}
 \acro{IR}{incremental redundancy}
 \acro{IRC}{interference rejection combining}
 \acro{ISD}{inter site distance}
 \acro{ISI}{inter symbol interference}
 \acro{ITU}{International Telecommunication Union}
 \acro{Iub}{UMTS interface between RNC and NodeB}
 \acro{IWF}{IMS interworking function}
 \acro{JFI}{Jain's fairness index}
 \acro{KPI}{key performance indicator}
 \acro{KNN}{k-nearest neighbours}
 \acro{L1}{layer one}
 \acro{L2}{layer two}
 \acro{L3}{layer three}
 \acro{LA}{location area}
 \acro{LAA}{licensed Assisted Access}
 \acro{LAC}{location area code}
 \acro{LAI}{location area identity}
 \acro{LAU}{location area update}
 \acro{LDA}{linear discriminant analysis} 
 \acro{LIDAR}{laser imaging detection and ranging}
 \acro{LLR}{log-likelihood ratio}
 \acro{LLS}{link-level simulation}
 \acro{LMDS}{local multipoint distribution service}
 \acro{LMMSE}{linear minimum mean-square-error}
 \acro{LoS}{line-of-sight}
 \acro{LPC}{logical PDCCH candidate}
 \acro{LPN}{low power node}
 \acro{LR}{likelihood ratio}
 \acro{LSAS}{large-scale antenna system}
 \acro{LSP}{large-scale parameter}
 \acro{LSTM}{long short term memory cell}
 \acro{LTE/SAE}{long term evolution/system architecture evolution}
 \acro{LTE}{long term evolution}
 \acro{LTE-A}{long term evolution advanced}
 \acro{LUT}{look up table}
 \acro{MAC}{medium access control}
 \acro{MaCe}{macro cell}
  \acro{MAE}{mean absolute error}
 \acro{MAP}{media access protocol}
 \acro{MAPE}{mean absolute percentage error}
 \acro{MAXI}{maximum insertion}
 \acro{MAXR}{maximum removal}
 \acro{MBMS}{multicast broadcast multimedia service} 
 \acro{MBS}{macrocell base station}
 \acro{MBSFN}{multicast-broadcast single-frequency network}
 \acro{MC}{modulation and coding}
 \acro{MCB}{main circuit board}
 \acro{MCM}{multi-carrier modulation}
 \acro{MCP}{multi-cell processing}
 \acro{MCS}{modulation and coding scheme}
 \acro{MCSR}{multi-carrier soft reuse}
 \acro{MDAF}{management data analytics function}
 \acro{MDP}{markov decision process }
 \acro{MDT}{minimisation of drive tests}
 \acro{MEA}{multi-element antenna}
 \acro{MeNodeB}{Master eNodeB}
 \acro{MGW}{media gateway}
 \acro{MIB}{master information block}
 \acro{MIC}{mean instantaneous capacity}
 \acro{MIESM}{mutual information effective SINR mapping}
 \acro{MIMO}{multiple-input multiple-output}
 \acro{MINI}{minimum insertion}
 \acro{MINR}{minimum removal}
 \acro{MIP}{mixed integer program}
 \acro{MISO}{multiple-input single-output}
 \acro{ML}{machine learning}
 \acro{MLB}{mobility load balancing}
 \acro{MLB}{mobility load balancing}
 \acro{MM}{mobility management}
 \acro{MME}{mobility management entity}
 \acro{mMIMO}{massive multiple-input multiple-output}
 \acro{MMSE}{minimum mean square error}
 \acro{mMTC}{massive machine type communication}
 \acro{MNC}{mobile network code}
 \acro{MNO}{mobile network operator}
 \acro{MOS}{mean opinion score}
 \acro{MPC}{multi-path component}
 \acro{MR}{measurement report}
 \acro{MRC}{maximal ratio combining}
 \acro{MR-FDPF}{multi-resolution frequency-domain parflow}
 \acro{MRO}{mobility robustness optimisation}
 \acro{MRT}{Maximum Ratio Transmission}
 \acro{MS}{mobile station}
 \acro{MSC}{mobile switching centre}
 \acro{MSE}{mean square error}
 \acro{MSISDN}{mobile subscriber integrated services digital network number}
 \acro{MUE}{macrocell user equipment}
 \acro{MU-MIMO}{multi-user MIMO}
 \acro{MVNO}{mobile virtual network operators}
 \acro{NACK}{negative acknowledgment}
 \acro{NAS}{non access stratum}
 \acro{NAV}{network allocation vector}
 \acro{NB}{Naive Bayes}   
 \acro{NCL}{neighbour cell list}
 \acro{NEE}{network energy efficiency}
  \acro{NF}{Network Function}
 \acro{NFV}{Network Functions Virtualization}
 \acro{NG}{next generation}
 \acro{NGMN}{next generation mobile networks}
 \acro{NG-RAN}{next generation radio access network} 
 \acro{NIR}{non ionisation radiation}
 \acro{NLoS}{non-line-of-sight}
 \acro{NN}{nearest neighbour} 
 \acro{NR}{new radio}
 \acro{NRTPS}{non-real-time polling service}
 \acro{NSS}{network switching subsystem}
 \acro{NTP}{network time protocol}
 \acro{NWG}{network working group}
 \acro{NWDAF}{network data analytics function} 
 \acro{OA}{open access}
 \acro{OAM}{operation, administration and maintenance}
 \acro{OC}{optimum combining}
 \acro{OCXO}{oven controlled oscillator}
 \acro{ODA}{omdi-directional antenna} 
 \acro{ODU}{optical distribution unit}
 \acro{OFDM}{orthogonal frequency division multiplexing}
 \acro{OFDMA}{orthogonal frequency division multiple access}
 \acro{OFS}{orthogonally-filled subframe}
 \acro{OLT}{optical line termination}
 \acro{ONT}{optical network terminal}
 \acro{OPEX}{operational expenditure}
 \acro{OSF}{odd subframe}
 \acro{OSI}{open systems interconnection}
 \acro{OSS}{operation support subsystem}
 \acro{OTT}{over the top}
 \acro{P2MP}{point to multi-point}
 \acro{P2P}{point to point}
 \acro{PAPR}{peak-to-average power ratio}
 \acro{PA}{power amplifier}
 \acro{PBCH}{physical broadcast channel}
 \acro{PC}{power control}
 \acro{PCB}{printed circuit board}
 \acro{PCC}{primary carrier component}
 \acro{PCCH}{paging control channel}
 \acro{PCCPCH}{primary common control physical channel}
 \acro{PCell}{primary cell}
 \acro{PCFICH}{physical control format indicator channel}
 \acro{PCH}{paging channel}
 \acro{PCI}{physical layer cell identity}
 \acro{PCPICH}{primary common pilot channel}
 \acro{PCPPH}{physical common packet channel}
 \acro{PDCCH}{physical downlink control channel}
 \acro{PDCP}{packet data convergence protocol}
 \acro{PDF}{probability density function}
 \acro{PDSCH}{physical downlink shared channel}
 \acro{PDU}{packet data unit}
 \acro{PeNB}{pico eNodeB}
 \acro{PeNodeB}{pico eNodeB}
 \acro{PF}{proportional fair}
 \acro{PGW}{packet data network gateway}
 \acro{PGFL}{probability generating functional}
 \acro{PhD}{doctor of philosophy}
 \acro{PHICH}{physical HARQ indicator channel}
 \acro{PHY}{physical}
 \acro{PIC}{parallel interference cancellation}
 \acro{PKI}{public key infrastructure}
 \acro{PL}{path loss}
 \acro{PMI}{precoding-matrix indicator}
 \acro{PLMN ID}{public land mobile network identity}
 \acro{PLMN}{public land mobile network}
 \acro{PML}{perfectly matched layer}
 \acro{PMF}{probability mass function}
 \acro{PMP}{point to multi-point}
 \acro{PN}{pseudorandom noise}
 \acro{POI}{point of interest}
 \acro{PON}{passive optical network}
 \acro{POP}{point of presence}
 \acro{PP}{point process}
 \acro{PPP}{Poisson point process}
 \acro{PPT}{PCI planning tools}
 \acro{PRACH}{physical random access channel}
 \acro{PRB}{physical resource block}
 \acro{PSC}{primary scrambling code}
 \acro{PSD}{power spectral density}
 \acro{PSS}{primary synchronisation channel}
 \acro{PSTN}{public switched telephone network}
 \acro{PTP}{point to point}
 \acro{PUCCH}{Physical Uplink Control Channel}
 \acro{PUE}{picocell user equipment}
 \acro{PUSC}{partial usage of subchannels}
 \acro{PUSCH}{physical uplink shared channel}
 \acro{QAM}{quadrature amplitude modulation}
 \acro{QCI}{QoS class identifier}
 \acro{QoE}{quality of experience}
 \acro{QoS}{quality of service}
 \acro{QPSK}{quadrature phase-shift keying}
 \acro{RAB}{radio access bearer}
 \acro{RACH}{random access channel}
 \acro{RADIUS}{remote authentication dial-in user services}
 \acro{RAN}{radio access network}
 \acro{RANAP}{radio access network application part}
 \acro{RAT}{radio access technology}
 \acro{RAU}{remote antenna unit}
 \acro{RAXN}{relay-aided x network}
 \acro{RB}{resource block}
 \acro{RCI}{re-establish cell id}
 \acro{RE}{resource efficiency}
 \acro{REB}{range expansion bias}
 \acro{REG}{resource element group}
 \acro{RF}{radio frequency}
  \acro{RFID}{radio frequency identification}
 \acro{RFP}{radio frequency planning}
 \acro{RI}{rank indicator}
 \acro{RL}{reinforcement learning}
 \acro{RLC}{radio link control}
 \acro{RLF}{radio link failure}
 \acro{RLM}{radio link monitoring}
 \acro{RMA}{rural macrocell}
 \acro{RMS}{root mean square}
 \acro{RMSE}{root mean square error}
 \acro{RN}{relay node}
 \acro{RNC}{radio network controller}
 \acro{RNL}{radio network layer}
 \acro{RNN}{recurrent neural network}
 \acro{RNP}{radio network planning}
 \acro{RNS}{radio network subsystem}
 \acro{RNTI}{radio network temporary identifier}
 \acro{RNTP}{relative narrowband transmit power}
 \acro{RPLMN}{registered PLMN}
 \acro{RPSF}{reduced-power subframes}
 \acro{RR}{round robin}
 \acro{RRC}{radio resource control}
 \acro{RRH}{remote radio head}
 \acro{RRM}{radio resource management}
 \acro{RS}{reference signal}
 \acro{RSC}{recursive systematic convolutional}
 \acro{RS-CS}{resource-specific cell-selection}
 \acro{RSQ}{reference signal quality}
 \acro{RSRP}{reference signal received power}
 \acro{RSRQ}{reference signal received quality}
 \acro{RSS}{reference signal strength}
 \acro{RSSI}{receive signal strength indicator}
 \acro{RTP}{real time transport}
 \acro{RTPS}{real-time polling service}
 \acro{RTS}{request-to-send}
 \acro{RTT}{round trip time}
  \acro{RU}{remote unit}
  \acro{RV}{random variable}
 \acro{RX}{receive}
 \acro{S1-AP}{s1 application protocol}
 \acro{S1-MME}{s1 for the control plane}
 \acro{S1-U}{s1 for the user plane}
 \acro{SA}{simulated annealing}
 \acro{SACCH}{slow associated control channel}
 \acro{SAE}{system architecture evolution}
 \acro{SAEGW}{system architecture evolution gateway}
 \acro{SAIC}{single antenna interference cancellation}
 \acro{SAP}{service access point}
 \acro{SAR}{specific absorption rate}
 \acro{SARIMA}{seasonal autoregressive integrated moving average}
 \acro{SAS}{spectrum allocation server}
 \acro{SBS}{super base station}
 \acro{SCC}{standards coordinating committee}
 \acro{SCCPCH}{secondary common control physical channel}
 \acro{SCell}{secondary cell}
 \acro{SCFDMA}{single carrier FDMA}
 \acro{SCH}{synchronisation channel}
 \acro{SCM}{spatial channel model}
 \acro{SCN}{small cell network}
 \acro{SCOFDM}{single carrier orthogonal frequency division multiplexing}
 \acro{SCP}{single cell processing}
 \acro{SCTP}{stream control transmission protocol}
 \acro{SDCCH}{standalone dedicated control channel}
 \acro{SDMA}{space-division multiple-access}
  \acro{SDO}{standard development organization}
 \acro{SDR}{software defined radio}
 \acro{SDU}{service data unit}
 \acro{SE}{spectral efficiency}
 \acro{SeNodeB}{secondary eNodeB}
 \acro{Seq2Seq}{Sequence-to-sequence}
 \acro{SFBC}{space frequency block coding}
 \acro{SFID}{service flow ID}
 \acro{SG}{signalling gateway}
 \acro{SGSN}{serving GPRS support node}
 \acro{SGW}{serving gateway}
 \acro{SI}{system information}
 \acro{SIB}{system information block}
 \acro{SIB1}{systeminformationblocktype1}
 \acro{SIB4}{systeminformationblocktype4}
 \acro{SIC}{successive interference cancellation}
 \acro{SIGTRAN}{signalling transport}
 \acro{SIM}{subscriber identity module}
 \acro{SIMO}{single input multiple output}
 \acro{SINR}{signal to interference plus noise ratio}
 \acro{SIP}{session initiated protocol}
 \acro{SIR}{signal to interference ratio}
 \acro{SISO}{single input single output}
 \acro{SLAC}{stochastic local area channel}
 \acro{SLL}{secondary lobe level}
 \acro{SLNR}{signal to leakage interference and noise ratio}
 \acro{SLS}{system-level simulation}
 \acro{SMAPE}{symmetric mean absolute percentage error}
 \acro{SMB}{small and medium-sized businesses}
 \acro{SmCe}{small cell}
 \acro{SMS}{short message service}
 \acro{SN}{serial number}
 \acro{SNMP}{simple network management protocol}
 \acro{SNR}{signal to noise ratio}
 \acro{SOCP}{second-order cone programming}
 \acro{SOHO}{small office/home office}
 \acro{SON}{self-organising network}
 \acro{son}{self-organising networks}
 \acro{SOT}{saving of transmissions}
 \acro{SPS}{spectrum policy server}
 \acro{SRS}{sounding reference signals}
 \acro{SS}{synchronization signal}
 \acro{SSL}{secure socket layer}
 \acro{SSMA}{spread spectrum multiple access}
 \acro{SSS}{secondary synchronisation channel}
 \acro{ST}{spatio temporal}
 \acro{STA}{steepest ascent}
 \acro{STBC}{space-time block coding}
 \acro{SUI}{stanford university interim}
 \acro{SVR}{support vector regression}
 \acro{TA}{timing advance}
 \acro{TAC}{tracking area code}
 \acro{TAI}{tracking area identity}
 \acro{TAS}{transmit antenna selection}
 \acro{TAU}{tracking area update}
 \acro{TCH}{traffic channel}
 \acro{TCO}{total cost of ownership}
 \acro{TCP}{transmission control protocol}
 \acro{TCXO}{temperature controlled oscillator}
 \acro{TD}{temporal difference}
 \acro{TDD}{time division duplexing}
 \acro{TDM}{time division multiplexing}
 \acro{TDMA}{time division multiple access}
  \acro{TDoA}{time difference of arrival}
 \acro{TEID}{tunnel endpoint identifier}
 \acro{TLS}{transport layer security}
 \acro{TNL}{transport network layer}
  \acro{ToA}{time of arrival}
 \acro{TP}{throughput}
 \acro{TPC}{transmit power control}
 \acro{TPM}{trusted platform module}
 \acro{TR}{transition region}
 \acro{TS}{tabu search}
 \acro{TSG}{technical specification group}
 \acro{TTG}{transmit/receive transition gap}
 \acro{TTI}{transmission time interval}
 \acro{TTT}{time-to-trigger}
 \acro{TU}{typical urban}
 \acro{TV}{television}
 \acro{TWXN}{two-way exchange network}
 \acro{TX}{transmit}
 \acro{UARFCN}{UTRA absolute radio frequency channel number}
 \acro{UAV}{unmanned aerial vehicle}
 \acro{UCI}{uplink control information}
 \acro{UDP}{user datagram protocol}
 \acro{UDN}{ultra-dense network}
 \acro{UE}{user equipment}
 \acro{UGS}{unsolicited grant service}
 \acro{UICC}{universal integrated circuit card}
 \acro{UK}{united kingdom}
 \acro{UL}{uplink}
 \acro{UMA}{unlicensed mobile access}
 \acro{UMi}{urban micro}
 \acro{UMTS}{universal mobile telecommunication system}
 \acro{UN}{United Nations}
 \acro{URLLC}{ultra-reliable low-latency communication}
 \acro{US}{upstream}
 \acro{USIM}{universal subscriber identity module}
 \acro{UTD}{theory of diffraction}
 \acro{UTRA}{UMTS terrestrial radio access}
 \acro{UTRAN}{UMTS terrestrial radio access network}
 \acro{UWB}{ultra wide band}
 \acro{VD}{vertical diffraction}
 \acro{VDFP}{vertical dynamic frequency planning}
 \acro{VDSL}{very-high-bit-rate digital subscriber line}
 \acro{VeNB}{virtual eNB}
 \acro{VeNodeB}{virtual eNodeB}
 \acro{VIC}{victim cell}
 \acro{VLR}{visitor location register}
 \acro{VNF}{virtual network function}
 \acro{VoIP}{voice over IP}
 \acro{VoLTE}{voice over LTE}
 \acro{VPLMN}{visited PLMN}
 \acro{VR}{visibility region}
  \acro{VRAN}{virtualized radio access network}
 \acro{WCDMA}{wideband code division multiple access}
 \acro{WEP}{wired equivalent privacy}
 \acro{WG}{working group}
 \acro{WHO}{world health organisation}
 \acro{Wi-Fi}{Wi-Fi}
 \acro{WiMAX}{wireless interoperability for microwave access}
 \acro{WiSE}{wireless system engineering}
 \acro{WLAN}{wireless local area network}
 \acro{WMAN}{wireless metropolitan area network}
 \acro{WNC}{wireless network coding}
 \acro{WRAN}{wireless regional area network}
 \acro{WSEE}{weighted sum of the energy efficiencies}
 \acro{WPEE}{weighted product of the energy efficiencies}
 \acro{WMEE}{weighted minimum of the energy efficiencies}
 \acro{X2}{x2}
 \acro{X2-AP}{x2 application protocol}
 \acro{ZF}{zero forcing}

 \end{acronym}

\end{document}